\documentstyle[prd,aps,epsf]{revtex}
\tighten
\begin{document}
\draft

\preprint{cond-mat/??? Imperial/TP/??}

\twocolumn[\hsize\textwidth\columnwidth\hsize\csname @twocolumnfalse\endcsname

\title{Non-Equilibrium Duality}
\author{R.\ J.\ Rivers}
\address{Blackett Laboratory, Imperial College, London SW7 2BZ}
\date{\today}
\maketitle

\begin{abstract}
The dual 'worldline', or 'string', description of adiabatic Ginzburg-Landau field theory
near a phase transition in terms of quasi-Brownian strings and loops is
well understood. In reality, the implementation of a transition is
intrinsically non-equilibrium. We sketch how time-dependent
Ginzburg-Landau theory leads to a modified dual string picture in
which  a causal bound on the growth of unstable string prevents
the uncontrolled proliferation of string (the Shockley-Hagedorn
transition) suggested by the adiabatic approximation.

\end{abstract}


\vskip2pc]

From the early days of lattice
models\cite{Stone} it has been appreciated that
time-independent Ginzburg-Landau (TIGL) theory has a dual
'worldline' representation in terms of near-Brownian strings and loops.  More
details are given in Refs. \cite{Kleinert,Schakel}.

The main ingredient in all these calculations is that they are
performed for systems in {\it equilibrium}, in which the
temperature $T$ is fixed at values close to the critical
temperature $T_c$. For the case that interests us, that of
continuous transitions, the instabilities that characterise the
transitions are, in this adiabatic dual picture, a consequence of
the uncontrolled proliferation of string.

In practice, transitions occur  in a finite, often short, time in
which the temperature $T$ crosses $T_c$ from its initial value.
The adiabatic approximation is only valid away from the
transition, when the field can keep in step with the changing
environment. Close to the transition the adiabatic regime is
replaced by an impulse regime, in which the field falls out of
step with what its equilibrium behaviour would be, and freezes
in\cite{zurek1,zurek2}. Just as the correlation length, in
reality, is unable to diverge because there is not enough time, so
we would expect a realistic dual worldline description in terms of strings
to show a limit to their proliferation as we pass through a
transition.

The question that we shall begin to address here is how the intrinsically
{\it non-adiabatic} transition has a dual representation in terms of
strings and loops.

However, we begin with a brief recapitulation of equilibrium dual theory.
The simplest condensed matter TIGL theory is
that of a single complex field $\phi$,
with Ginzburg-Landau
free energy
\[
F(T) = \int d^{3}x\,\,\bigg(\frac{\hbar^{2}}{2m}|\nabla\phi |^{2}
+\alpha (T)|\phi |^{2} + \beta |\phi |^{4}\bigg)
\]
in which the chemical potential $\alpha (T) =
\alpha_{0}\epsilon(T)$, where $\epsilon (T) = (T/T_c -1)$,
vanishes at $T_{c}$.  Such an energy provides a reasonable
description of superfluid $^4He$, a simplified model for $^3 He$
and a good description of the scalar sector of low-$T_c$
superconductors.

It is convenient to work in spatial units of $\xi_0
=\sqrt{\hbar^2/2m\alpha_0}$ when, on rescaling the field,
self-coupling and temperature,
\[
{\bar F}(T) = \int d^{3}x\,\,\bigg(|\nabla\phi |^{2} + \epsilon
(T)|\phi |^{2} + {\bar\beta} |\phi |^{4}\bigg).
\]
At temperature $T_0 > T_c$ the {\it free-field} correlation function
that follows from
\[
{\bar F}_0 (T_0) = \int d^{3}x\,\,\bigg(|\nabla\phi |^{2} +
\epsilon (T_0)|\phi |^{2}\bigg)
\]
is
\[
\langle\phi ({\bf x})\phi^* ({\bf 0})\rangle = G_0 (r) = \int d \!
\! \! / ^3 k \, e^{i {\bf k} . {\bf x} } P(k),
\]
($r = |{\bf x}|$) in which the power spectrum
\[
P(k) =\frac{1}{{\bf k}^2 + \epsilon (T_0)} = \int_{0}^{\infty}
d\tau \,e^{-\tau k^{2}}\,e^{-\tau\epsilon (T_0)}
\]
has the usual representation in terms of the Schwinger proper-time
$\tau$. In turn, this gives
\begin{equation}
G_0 (r) = \int_{0}^{\infty} d\tau\,
\bigg(\frac{1}{4\pi\tau}\bigg)^{3/2}
e^{-r^{2}/4\tau}\,e^{-\tau\epsilon (T_0)}. \label{lgcorr}
\end{equation}
$G_0 (0)$, necessary for loops, diverges
 from the UV singularities at $\tau = 0$, but these can be
regulated with a fixed cutoff. We choose a cutoff of order unity
in units of $\xi_0 $, which will be implicit throughout.

The dual picture is obtained by observing that $(1/4\pi\tau
)^{3/2} e^{-r^{2}/4\tau}$ is the probability distribution for a
Brownian 'worldline' or, more usefully, a 'polymer', path ${\bf x}(\tau)$,
parametrised by $\tau$,  having its endpoints separated by a
distance $r$. Specifically, it is expressible as the sum over such
paths
\[
\bigg(\frac{1}{4\pi\tau}\bigg)^{3/2}
e^{-r^{2}/4\tau}=
\int^{{\bf x}(\tau)={\bf x}}_{{\bf x}(0)={\bf 0}}
{\cal D}{\bf x}\,\exp\bigg[-\int_{0}^{\tau} d\tau '\,\,
\frac{1}{4}(\frac{d{\bf x}}{d\tau '})^{2}\bigg].
\]
If we think of the path as having step length unity (in units of
$\xi_0$) then $\tau$ is proportional to the length of the
path\cite{Schakel} and we shall use $\tau$ and path length
synonymously.

Then $G_0 (r)$ is the sum over string paths of all lengths,
\[
G_0 (r)
=\int_{0}^{\infty} d\tau\,
\int^{{\bf x}(\tau)={\bf x}}_{{\bf x}(0)={\bf 0}}
{\cal D}{\bf x}\,e^{- S_{eq} [{\bf x};\tau, \epsilon (T_0)]},
\]
where $S_{eq} [{\bf x};\tau, \epsilon (T_0)]$ is the equilibrium Euclidean action
\[
S_{eq} [{\bf x};\tau, \epsilon (T_0)]=\int_{0}^{\tau} d\tau '\,\,
\bigg(\frac{1}{4}(\frac{d{\bf x}}{d\tau '})^{2} +\epsilon (T_0) \bigg).
\]

The free-field partition function
\[
{\cal Z} = \int{\cal D}\phi\,{\cal D}\phi^*\,\,e^{- {\bar F}_0} =
\exp [-tr\ln (-\nabla^{2} + \epsilon (T_0))]
\]
is
just that of a gas of free orientable polymer or string {\it loops},
whose lengths are parametrised by $\tau$, with one-loop partition function
\begin{equation}
\ln{\cal Z} =
\int_{0}^{\infty} \frac{d\tau}{\tau}\,
\bigg(\frac{1}{4\pi\tau}\bigg)^{3/2} \,e^{-\tau\epsilon (T_0)}.
\label{length}
\end{equation}
$\epsilon (T_0)$ is
understood as the energy/unit length or tension of the string.
The factor $\propto \tau^{-5/2}$ in (\ref{length}) is the length distribution
for Brownian strings of length $\tau$. As
we drive $T_0\rightarrow T_c\,+$ it
is easier to create string. If, formally, we take $T_0$ below $T_c$
then the onset of '{\it negative}' tension makes long loops overwhelmingly
favoured and the condensation of loops that
follows is understood by condensed matter physicists as a Shockley-Feynman
transition, and by quantum field theorists as a Hagedorn
transition.

The
average loop length is
\begin{equation}
\langle\tau\rangle = -\frac{d\,\ln{\cal Z}}{d\,\epsilon (T_0)}
 = \frac{\int_{0}^{\infty} d\tau\,
\bigg(\frac{1}{4\pi\tau}\bigg)^{3/2} \,e^{-\tau\epsilon (T_0)}
}{\int_{0}^{\infty} \frac{d\tau}{\tau}\,
\bigg(\frac{1}{4\pi\tau}\bigg)^{3/2} \,e^{-\tau\epsilon (T_0)} }.
\label{ll}
\end{equation}
Yet again, forcing $T$ below $T_c$ makes $\langle\tau\rangle$
diverge from the IR divergence of the integrals at large $\tau$. A
formal IR cutoff at $\tau = \tau_{max}$ leads to $\langle\tau\rangle
= O(\tau_{max})$.

 The incorporation of the ${\bar\beta}|\phi|^{4}$ field
self-interaction into equilibrium dual string theory is
implemented by the introduction of a repulsive steric
string interaction at the points where paths or strings
cross\cite{Kleinert}. Although this reduces the weight of string
configurations, qualitatively the transition still occurs in this
adiabatic picture because of the  proliferation of strings.

In practice, a change of phase is enforced by changing the sign of
$\epsilon (t)$, either by reducing $T$, or by varying the critical
temperature $T_c$. Experiments with superconductors and (the
neutron bombardment of) $^3 He$  do the former, while pressure
quenches of $^4 He$ do the latter. If we write
\[
\epsilon (t) = \epsilon (T(t)) =\frac{T(t)}{T_c(t)} - 1,
\]
the {\it adiabatic approximation} for the 'free' correlation function $G_0 (r, t)$ at
time $t$ is
\begin{equation}
G^{ad}_0 (r,t)=\int_{0}^{\infty} d\tau\, \int^{{\bf x}(\tau)={\bf
x}}_{{\bf x}(0)={\bf 0}} {\cal D}{\bf
x}\,e^{-S_{eq}[{\bf x}, \tau ,\epsilon (t)]},
\label{lgcorr6}
\end{equation}
in which we just make a straightforward substitution of the
equilibrium $\epsilon (T_0 )$ with $\epsilon (t)$ in $S_{eq}[{\bf x},\tau, \epsilon (T_0 )]$. This is
treating the equilibrium pictures as a series of snapshots that
can be run together as a continuous film. In this film string lengths
increase uncontrollably as we cross the transition.
In particular, $G^{ad}_{0}(r,t)$
of (\ref{lgcorr}) is simply calculable as the Yukawa correlator
\[
G^{ad}_0 (r) = \frac{1}{4\pi r}\,e^{-r/\xi_{ad} (t)},
\]
where, on rescaling, $\xi_{ad} (t) = \xi_0 / \sqrt{\epsilon(t )}$
 diverges as $T(t)\rightarrow T_c$.
 This cannot be the case, since causality alone prevents the
correlation length diverging in a finite time. In terms of the dual
picture this implies that the
production of an infinity of string in a finite time is equally
prohibited.  The question is, how?

To
answer this, we adopt the time-dependent Landau-Ginzburg (TDLG)
equation for ${\bar F}$,
\begin{equation}
\frac{1}{\Gamma}\frac{\partial\phi}{\partial t} = -\frac{\delta
{\bar F}}{\delta\phi} + \eta , \label{tdlg}
\end{equation}
where $\eta$ is Gaussian thermal noise, satisfying
\begin{equation}
\langle\eta ({\bf x},t)\eta^* ({\bf y},t')\rangle =
2T(t)\Gamma\delta ({\bf x}-{\bf y})\delta (t -t').
\label{noise}
\end{equation}

Let us continue to consider the free-field case, ${\bar F}={\bar
F}_0$. In (\ref{tdlg}) the natural unit of time is $\tau_0 =
1/\alpha_0\Gamma$ and, in units of $\tau_0$ and $\xi_0$,
Eq.(\ref{noise}) becomes
\begin{equation}
{\dot\phi}({\bf x},t) = - [-\nabla^{2} + \epsilon (t)]\phi
({\bf x},t) +{\bar\eta} ({\bf x},t). \label{free}
\end{equation}
where ${\bar\eta}$ is the renormalised noise. It is straightforward to show that the
equal-time correlation function is now
\[
\langle\phi ({\bf x},t)\phi^* ({\bf 0},t)\rangle =G_0 (r,t) = \int
d \! \! \! / ^3 k \, e^{i {\bf k} . {\bf x} } P(k, t).
\]
in which the power spectrum $P(k,t)$ has a representation in terms
of the Schwinger proper-time $\tau$ as
\[
P(k, t) = \int_{0}^{\infty} d\tau \,{\bar T}(t-\tau/2)\,e^{-\tau
k^{2}}\,e^{-\int_{0}^{\tau} d\tau'\,\,\epsilon (t- \tau'/2)},
\]
where ${\bar T}$ is the renormalised temperature. In turn, this
gives

\noindent $G_0 (r, t) = $
\begin{equation}
\int_{0}^{\infty} d\tau\,{\bar T}(t-\tau/2)
\,\bigg(\frac{1}{4\pi\tau}\bigg)^{3/2}
e^{-r^{2}/4\tau}\,e^{-\int_{0}^{\tau} d\tau'\,\,\epsilon (t- \tau'/2)}.
\label{lgcorr7}
\end{equation}
This differs significantly from (\ref{lgcorr6}), which would
replace $\epsilon (t - \tau'/2)$ in (\ref{lgcorr7})with $\epsilon (t)$. Further,
because of ${\bar T}(t-\tau/2)$, the paths are no longer Brownian
in their length distribution.

However, $G_0 (r,t)$ can still be expressed in terms of paths, as
\[
G_0 (r,t) =\int_{0}^{\infty} d\tau\,{\bar T}(t-\tau/2) \int^{{\bf
x}(\tau)={\bf x}}_{{\bf x}(0)={\bf 0}} {\cal D}{\bf
x}\,e^{-S[{\bf x},\tau, t]}.
\]
$S[{\bf x},\tau, t]$ is not the equilibrium action
$S_{eq}[{\bf x},\tau, \epsilon (t)]$, but
\[
S[{\bf x},\tau, t]=\int_{0}^{\tau} d\tau '\,\,
\bigg(\frac{1}{4}(\frac{d{\bf x}}{d\tau '})^{2} +\epsilon (t -\tau
'/2) \bigg).
\]
Unlike the case for the adiabatic approximation (\ref{lgcorr6}), the local tension
$\epsilon (t -\tau /2)$ is dependent on {\it both} the time $t$ at which
we examine the strings and on the position $\tau$ along them.
As a result the integrated tension
$\int_0^\tau d\tau'\epsilon (t - \tau'/2)$ does {\it not}
vanish at the transition time. Since it is the uniform vanishing of
tension which triggers the avalanche of string production we see
already that it will not happen in this case.

As a simple, if unrealistic, demonstration that transitions
implemented in a finite time do not impose singular dual behaviour we
consider the case of an {\it instantaneous} quench at time $t=0$
from a temperature $T_0$ above the critical temperature $T_c$ to
absolute zero. That is, $\epsilon (t) = \epsilon (T_0 )\theta (-t)$.

Simple calculation shows that  $G_0 (r,t)$ takes the form of (\ref{lgcorr})
for $t < 0$, whereas for $t > 0$
\begin{eqnarray}
G_0 (r,t) &=& e^{2tT_0/T_c}\int_{2t}^{\infty} d\tau\,
 \int^{{\bf x}(\tau)={\bf x}}_{{\bf x}(0)={\bf 0}} {\cal
D}{\bf x}\,e^{-S_{eq}[{\bf x},\tau, \epsilon (T_0) ]}
\\
\nonumber &=& e^{2tT_0/T_c} \int_{2t}^{\infty} d\tau\,
\bigg(\frac{1}{4\pi\tau}\bigg)^{3/2}
e^{-r^{2}/4\tau}\,e^{-\tau\epsilon (T_0)}. \label{lgcorrt}
\end{eqnarray}
After the transition we have a representation in terms of positive tension
 paths at the {\it initial} temperature $T_0$. The unstable (negative tension)
 paths, for which $\tau < 2t$,
 are {\it totally excluded}. Instead, the
instabilities are encoded in the non-singular exponential prefactor.

This lack of
IR singular behaviour is made even more explicit in a saddle-point
approximation for $G_0 (r,t)$. For $r^2
> 4\epsilon (T_0)t$ we find
\[
G_0 (r,t) \sim e^{2tT_0/T_c} \frac{1}{4\pi r}\,e^{-r/\xi},
\]
where $\xi =\xi (T_0) $ remains frozen in at its initial
equilibrium value. Unlike $\xi_{ad}(t)$ there is no divergence of
$\xi$ as we cross $T=T_c$ in this abrupt way.

For the equilibrium theory, $\langle\tau\rangle$ of (\ref{ll}) can
also be expressed as
\[
\langle\tau\rangle = \frac{G_0 (0)}{2G_0 ''(0)},
\]
where the prime denotes differentiation with respect to $r$. If we
adopt the same definition out of equilibrium the benign effect of
the quench to $T_f = 0$ is again apparent in that
$\langle\tau\rangle_t$ agrees with (\ref{ll}) for $t<0$, but is
\begin{equation}
\langle\tau\rangle_{t}
 = \frac{\int_{2t}^{\infty} d\tau\,
\bigg(\frac{1}{4\pi\tau}\bigg)^{3/2} \,e^{-\tau\epsilon (T_0)}
}{\int_{2t}^{\infty} \frac{d\tau}{\tau}\,
\bigg(\frac{1}{4\pi\tau}\bigg)^{3/2} \,e^{-\tau\epsilon (T_0)} }
\label{nell}
\end{equation}
for $t>0$. The exponential prefactors have cancelled to reproduce
the equilibrium result (\ref{ll}), again at the initial
temperature $T=T_0$, but for the absence of unstable loops with
length less than $2t$, to which there is now no reference. Because
of the exponential damping of long string the dominant string length is $\tau = 2t$.

 There is no
question of an IR divergence of loop length as naively suggested
by the equilibrium theory. We understand the stability of loops with length $\tau >2t$
in (\ref{nell}) as a causal bound. In our units, negative tension can only
propagate at speed $c=1$, which happens to be the cold speed of sound in the $\phi$ field.

This is reinforced by an extension to non-zero final temperature $T_f$
\begin{eqnarray}
\epsilon (t) &=& \epsilon (T_0) >0 \, \,\, \mbox{if $t<0$,}
\nonumber
\\
&=& \epsilon (T_f) < 0  \, \, \mbox{if $ t>0$}.
\nonumber
 \label{modes}
\end{eqnarray}
For $t > 0$ we now find
\begin{eqnarray}
G_0 (r,t) &=& e^{2t(T_0 - T_f )/T_c}\int_{2t}^{\infty} d\tau\,
 \int^{{\bf x}(\tau)={\bf x}}_{{\bf x}(0)={\bf 0}} {\cal
D}{\bf x}\,e^{-S_{eq}[{\bf x},\tau, \epsilon (T_0) ]}
\nonumber
\\
\nonumber
&+& \frac{T_f}{T_0}\int_{0}^{2t} d\tau\,
 \int^{{\bf x}(\tau)={\bf x}}_{{\bf x}(0)={\bf 0}} {\cal
D}{\bf x}\,e^{-S_{eq}[{\bf x},\tau, \epsilon (T_f) ]}.
\label{lgcorrtt}
\end{eqnarray}
We see explicitly how the limited length $\tau < 2t$ of string
with negative tension $\epsilon (T_f)< 0$ prevents such
string from giving a divergent contribution.
Small unstable loops are no longer precluded, but
their contribution is finite. The dominant length remains $\tau = 2t$.

Although an instantaneous quench is impossible, more general
quenches show similar
qualitative behaviour.
Specifically, suppose that $\epsilon (t)$ decreases
monotonically, with a single zero $\epsilon (0) = 0$ at time
$t=0$. As before, only strings of limited length $\tau < 2t$, for which $\epsilon (t
- \tau/2)<0$, have negative tension. However, for $\tau >2t$,
for which $\epsilon (t-\tau /2)>0$, strings have segments with negative and
positive tension. There is a causal bound $c=1$ on the speed at which
instability can propagate along a string.
The string with negative tension, with a contribution that is
independent of $\tau$, gives a prefactor growing at least exponentially.
The effect is to leave only stable string of
length $\tau >2t$ in the $\tau$ integral for $\tau >2t$.
This boundedness on unstable string prevents the
unlimited production of string suggested by the adiabatic
approximation.

The exponential growth of $G_0 (r,t)$ with time in  (\ref{lgcorrt}), and more generally,
can only be accommodated for a very short period, since $\langle
|\phi |^{2}\rangle = G_0 (0,t)$ must be constrained by the value
of the order parameter $\langle |\phi |\rangle$ after the
transition, equal to $\sqrt{{\bar\beta}^{-1}/2}$ in the absence of
corrections. A rough guide to the maximum time ${\bar t}$ for
which the free-field approximation is valid is that $G_{0}(0,{\bar
t})= {\bar\beta}^{-1}/2$.  As $t$ approaches ${\bar t}$, and
thereafter, the reaction of the field with itself will cut off the
exponential growth.

It is difficult to see how the dual string picture will survive at
later times without further approximation.  One indication is
through a mean-field (or large-N) approach in which the linear
nature of the theory is maintained self-consistently\cite{Boyanovsky}. In this
approach (\ref{free}) is replaced by
\begin{equation}
{\dot\phi}({\bf x},t) = - [-\nabla^{2} + \epsilon_{eff} (t)]\phi
({\bf x},t) +{\bar\eta} ({\bf x},t). \label{free2}
\end{equation}
with
\begin{eqnarray}
\epsilon_{eff}(t) &=& \epsilon (t) + p{\bar\beta}\langle |\phi
|^{2}\rangle
\nonumber
\\ &=& \epsilon (t) + p{\bar\beta}G(0,t),
\nonumber
\end{eqnarray}
where $G(0,t)$ is determined self-consistently from (\ref{free2}).
The coefficient $p$ depends on whether we adopt a mean-field
(Hartree) approximation or a large-N limit for $N=2$.

The assumed single zero of $\epsilon (t)$ at $t=0$ will lead to a
zero of $\epsilon_{eff}(t)$ at $t\approx 0$, and the previous
analysis applies; with only segments no longer than $2t$ with negative
tension, there is no singular behaviour as we cross the
transition.

To have a quantitative estimate  of the effect of back-reaction we
make a Gaussian approximation for $G(0,t)$ by expanding about the
zero of $\epsilon_{eff}(t)$. Assuming that the quench is not too
rapid, we find that, for $t\leq {\bar t}$, the average loop length is
\[
\langle\tau\rangle_{t}\approx
\frac{\int_{0}^{\infty}d\tau\,\frac{{\bar T}(t - \tau
/2)}{(4\pi\tau)^{3/2}} \,e^{-(\tau - 2t)^{2}|\epsilon '(0)|/4}}
{\int_{0}^{\infty} \frac{d\tau}{\tau}\,\frac{{\bar T}(t - \tau
/2)}{(4\pi\tau)^{3/2}} \,e^{-(\tau - 2t)^{2}|\epsilon '(0)|/4}},
\]
As before, the prefactors encoding the unstable strings cancel approximately,
and we need no information about the
self-consistent mass, but for the fact that $\epsilon_{eff} (t)\rightarrow 0$ at
large times  to stop
$G(0,t)$ from growing. We see the peak at $\tau = 2t$ in the length distribution moving clear
of the UV endpoint behaviour.
This continues for later times and $\langle\tau\rangle_t = O(t)$ once the
temperature has become low enough that ${\bar T}(t)$ has
suppressed the UV singular behaviour. Details are given in our
earlier work\cite{ray2,RKK}.

We stress that this discussion has been restricted to the 'first quantised' dual
representation of the Ginzburg-Landau theory.
However, it is a familiar result from a different viewpoint. For a linear system
like (\ref{free2}) it can be shown\cite{halperin,maz} that the
mean loop length $\langle\tau\rangle_{t}$ satisfies
\begin{equation}
\langle\tau\rangle_{t} = \frac{1}{4\pi n(t)}, \label{n(t)}
\end{equation}
where $n(t)$ is the density of {\it line zeroes} of the complex field
$\phi$.  The relevance of this is that, at later times, the global
{\it vortices} of this $U(1)$ scalar theory can be identified by
the line zeroes of their cores\cite{zurek3}. The linear growth of
{\it dual} loop lengths with time corresponds to a $t^{1/2}$ behaviour for
{\it $\phi$-field line-zero} separation and, when vortices are well-defined,
vortex separation\cite{zurek3}.
 This scaling behaviour can be
justified\cite{vinen} for the decay of vortices of $^{4}He$, and
is used to determine initial vortex densities at the $^{4}He$
transition\cite{lancaster}.

This work is the result of a network supported by the European Science
Foundation. I thank Adriaan Schakel for helpful comments.

\end{document}